\documentclass[
    reprint,
    aps,
    prl,
    floatfix
]{revtex4-2}

\usepackage{dcolumn}
\usepackage{bm}
\usepackage{xcolor}
\usepackage{graphicx}
\usepackage{svg}
\usepackage[percent]{overpic}
\usepackage{subcaption}

\begin{document}

\preprint{APS/123-QED}

\title{Ultralow Mean Transverse Energy and High Quantum Efficiency Cryogenic Bialkali Photocathode for MHz-repetition-rate Electron Sources }

\author{D. Wang}
\author{S. Liu}
\author{J. Liu}
\author{Y. Dai}
\author{Z. Hong}
\author{J. Wang}
\author{Y. Shi}
\author{M. Tai}
\author{L. Feng}
\author{H. Xu}
\author{L. Lin}
\author{F. Wang}
\author{F. Zhu}
\author{J. Hao}
\author{S. Quan}
\author{K. Liu}
\author{H. Xie}
\email{hmxie@pku.edu.cn}
\author{S. Huang}
\affiliation{State Key Laboratory of Nuclear Physics and Technology,Institute of Heavy Ion Physics, Peking University,
Beijing 100871, China}

\date{\today}

\begin{abstract}

Simultaneously achieving high quantum efficiency (QE), ultralow mean transverse energy (MTE), and robust long-term operation under conditions relevant to continuous-wave (CW) X-ray free-electron lasers (XFELs) remains a central challenge for semiconductor photocathodes. This challenge arises from the trade-off between QE and MTE, as well as the difficulty of maintaining stable operation in high-field CW electron guns. Here we demonstrate a cryogenic K$_2$CsSb photocathode that simultaneously achieves high QE, ultralow MTE, and robust long-term operation in a CW gun under XFEL-relevant operating conditions. Under cryogenic operation, the photocathode achieves an MTE of 50 meV while sustaining a QE of 5.4\%. Milliampere-level CW current, including operation at 5 mA, was demonstrated together with an approximately 20-day operational history. The observations are consistent with improved carrier survival and/or surface escape in photocathodes prepared using the optimized recipe. These results show that the practical QE-MTE trade-off can be substantially mitigated in cryogenic bialkali photocathodes, and provide a practical pathway toward high-brightness electron sources for CW XFELs and energy-recovery linacs.

\end{abstract}

\maketitle

High brightness electron beams are essential for a broad range of accelerator-based scientific facilities, including continuous-wave (CW) X-ray free-electron lasers (XFELs)~\cite{Petrushina2020HighBrightnessSRF,Sannibale2023HighDutyCycleXFEL}, energy recovery linacs (ERLs) \cite{Nakamura2015ERLFELDesign}, hadron coolers~\cite{Fedotov2020HadronCooling}, Compton gamma-ray sources\cite{Bacci2013ComptonGammaLinac}, and ultra-fast electron diffraction (UED)\cite{Maxson2017Sub10fsPRL}. The performance of these facilities is strongly determined by the initial phase-space distribution of electrons emitted from the photocathode. For photocathode-based electron sources, the achievable beam brightness is fundamentally governed by the photocathode quantum efficiency (QE), the mean transverse energy (MTE), the drive laser spot size, and the accelerating electric field at the cathode surface~\cite{Bazarov2009MaximumBrightness}.  The normalized beam brightness $B_n$can be expressed as
\begin{equation}
    B_n = \frac{2I}{\pi^2 \cdot \varepsilon_{n,x} \cdot \varepsilon_{n,y}}
    \label{eq:placeholder_label}
\end{equation}

where I is the beam current, $\varepsilon_{n,x} $ and $\varepsilon_{n,y}$ are transverse emittance. For fixed laser power, QE determines the extractable beam current, whereas MTE, together with the laser spot size, sets the cathode's intrinsic emittance. In low-emittance photoinjectors, this intrinsic contribution can constitute a substantial fraction of the total transverse emittance.

High QE is essential for realizing the practical benefit of an ultralow-MTE photocathode, because low QE requires higher laser power and can introduce additional thermal load, emission non-uniformity, and emittance growth. Thus, simultaneously achieving high QE and low MTE is crucial for translating the intrinsic photocathode advantage into a genuinely low-emittance, high-brightness beam in MHz-class CW electron sources.

\begin{figure}[!htbp]
    \centering
    \includegraphics[width=1\linewidth]{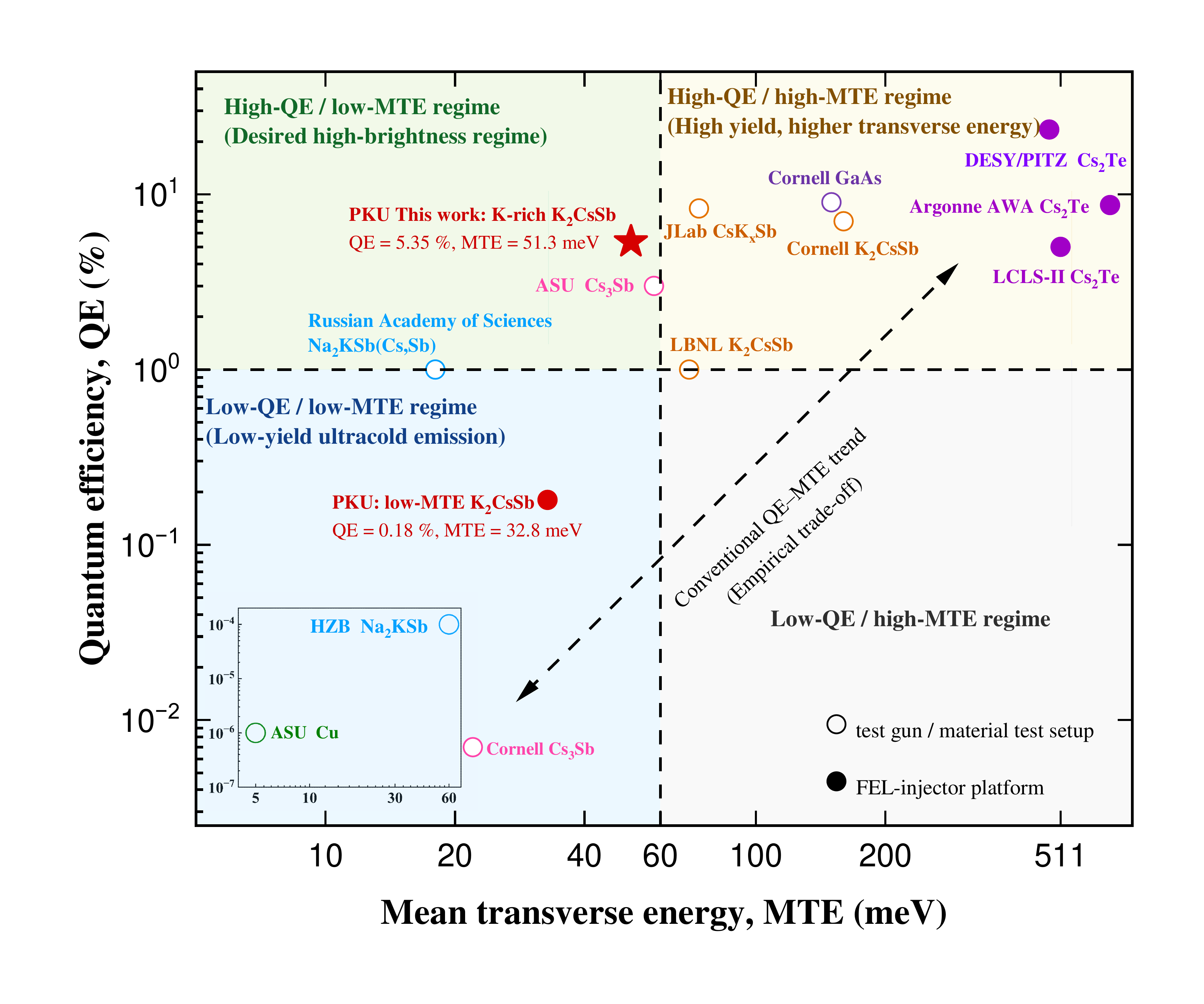} 
    \caption{QE–MTE performance regimes of representative photocathodes relevant to CW electron sources~\cite{Dube2025TripleEvaporation,Rozhkov2026Subthermal,Owusu2025Cs3SbCryogenic,Cultrera2015ColdElectronBeams,Karkare2020UltracoldCu,Bazarov2011CsK2SbThermalEmittance,Vecchione2011LowEmittance,Vecchione2012Roughness,HernandezGarcia2019CompactPhotogun,Cultrera2012CornellPhotocathodeRD,Huang2019PITZCs2Te,Zheng2020Cs2TeMapping,Zhou2023CWGuns,Zhou2021LCLSIICommissioning}.  }
    \label{fig:emittance}
\end{figure}

No published demonstration has yet combined percent-level QE, tens-of-meV MTE, and extended CW operation in an MHz-repetition-rate electron source relevant to XFELs. Cs$_2$Te photocathodes are widely used in XFEL injectors~\cite{JuarezLopez2024EuXFELPhotocathodes,Zhou2021LCLSIICommissioning,Zheng2023VHFGun} for their robustness and long operational lifetime, but their room-temperature mean transverse energy (MTE) typically exceeds 300~meV~\cite{Jones2024CERNCsTeMTE}. GaAs photocathodes operated in high-voltage dc guns generally exhibit lower MTE of approximately 90-150 meV\cite{Bazarov2008NEAGaAsMTE,Wu2015GaAsHVDCInjector}, but their limited operational lifetime under practical operating conditions remains a significant drawback\cite{Bae2022GaAsHVDCGunLifetime,Li2017IHEPDCGun}. Cryogenic operation has therefore been explored as a route to suppress the excess transverse energy. Liquid-nitrogen cooling of Cs$_2$Te at \cite{Xiang2010Cs2TeThermalEmittanceHZDR,Xiang2022AdvancedPhotocathodesSRFGuns} , however, produced little reduction in MTE.

Among semiconductor photocathodes, alkali antimonide, particularly K$_2$CsSb\cite{Gaowei2024PLDGrowth,Mohanty2023MultiAlkali} , are attractive for high-brightness electron sources because they combine percent-level QE under green-light illumination, sub-picosecond response, and relatively low intrinsic emittance. Long-lifetime operation of K$_2$CsSb photocathodes in 400 kV DC gun ~\cite{gunDunham2013CornellHighCurrent,Cultrera2011CornellHighCurrent,Gu2020LEReCDC} and a high-gradient continuous-wave SRF gun has also been demonstrated, establishing the practical viability of bialkali cathodes for high brightness CW injectors~\cite{Wang2021LongLifetimeSRF,Petrushina2020HighBrightnessSRF}. However, at room temperature their MTE remains typically 150--200~meV, corresponding to a normalized intrinsic emittance of approximately $0.50$-$0.56~\mathrm{mm\,mrad/mm}$~\cite{Bazarov2011CsK2SbThermalEmittance,Xu2024ThermalEmittanceBialkaliPKU}. Although cryogenic cooling can reduce the MTE to the tens-of-meV regime, it is commonly accompanied by a substantial loss of QE~\cite{Zhao2023QECryogenicBialkali,Xie2016CryogenicBialkaliPRAB,Maxson2013DIH}. Simultaneously preserving high QE, ultralow MTE, and practical operational stability in an XFEL-relevant electron gun therefore remains an important challenge.

Figure~\ref{fig:emittance} summarizes the QE--MTE performance landscape of representative photocathodes and electron-source platforms. An important route toward mitigating this QE--MTE trade-off was recently demonstrated by Arizona State University (ASU) using a fast-cooling strategy in a 200-kV cryogenic DC gun, with an MTE of approximately 58~meV and a QE of about 3\% from Cs$_3$Sb photocathodes ~\cite{Owusu2025Cs3SbCryogenic}. However, such performance has not yet been demonstrated in an MHz-repetition-rate electron source relevant to CW XFEL operation. Moreover, the relatively low electric field at the photocathode limits the achievable current density, making high-bunch-charge operation much more challenging. 

A similar challenge was encountered in the SRF lab at Peking University. The photocathodes prepared using our conventional sequential-deposition ("Conventional recipe") recipe~\cite{Ouyang2022BialkaliDCSRF} exhibited a significant degradation in performance after cooling to cryogenic temperature. The QE of the photocathode dropped to less than 20\% of its original value at room temperature($10^{-3}$ level). Cryogenic bialkali photocathodes routinely exhibited MTE in the 30--50~meV range after transfer and cooling(all "fast cooling process" as ASU\cite{Owusu2025Cs3SbCryogenic}). These observations suggest that achieving ultralow MTE itself is not the principal limitation; rather, the central challenge is to preserve efficient photoemission under the same cryogenic conditions. This motivated us to explore whether photocathode growth, rather than the cooling procedure alone, could provide an additional degree of freedom for recovering QE without sacrificing the ultralow MTE.

Here we show that the practical QE–MTE limitation\cite{Karkare2014UltrabrightGaAs} can be substantially mitigated by combining an optimized photocathode growth procedure with cryogenic operation in an MHz-repetition-rate electron source.   The following sections describe the photocathode preparation, and measurements of QE, MTE and operational stability. 

\begin{figure}[t]
    \centering

    \begin{subfigure}[t]{0.48\linewidth}
        \centering
           \begin{overpic}[
            width=\linewidth
        ]{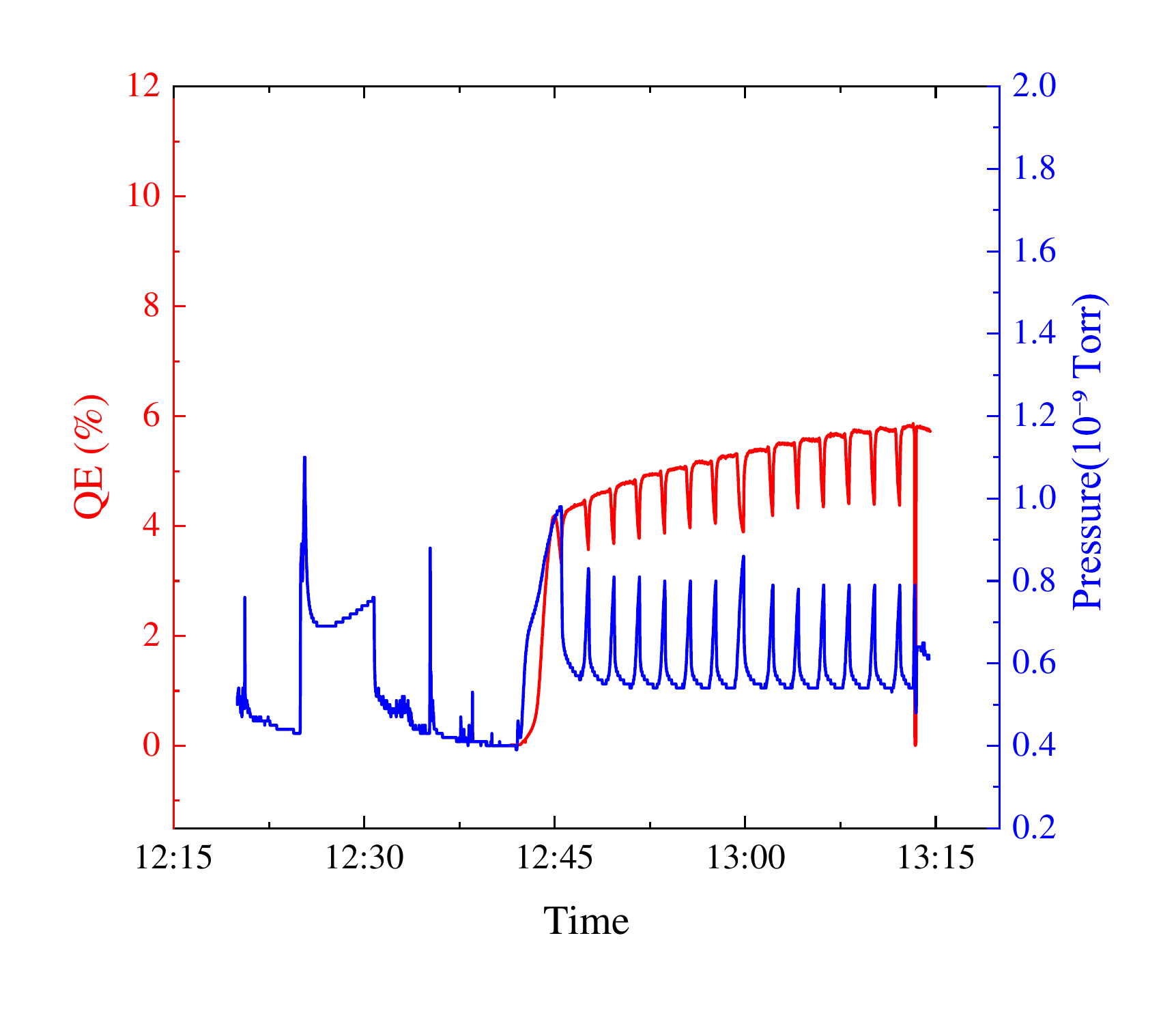}

            \put(17,46){%
                \includegraphics[
                    width=0.35\linewidth
                ]{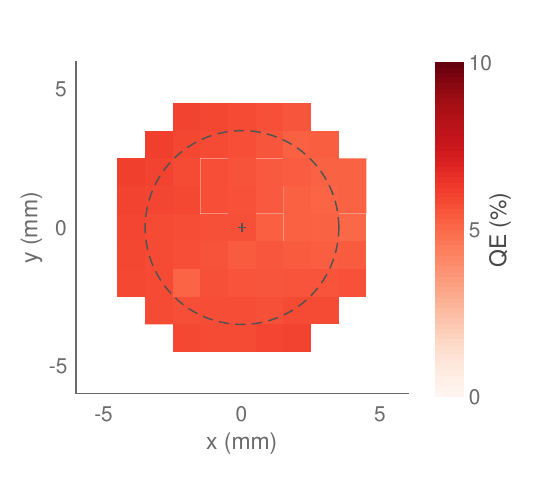}%
            }

        \end{overpic}      
        
        \caption{Conventional K$_2$CsSb photocathode.}
        \label{fig:qe_pressure_pk}
    \end{subfigure}
    \hfill
    \begin{subfigure}[t]{0.48\linewidth}
        \centering

        \begin{overpic}[
            width=\linewidth
        ]{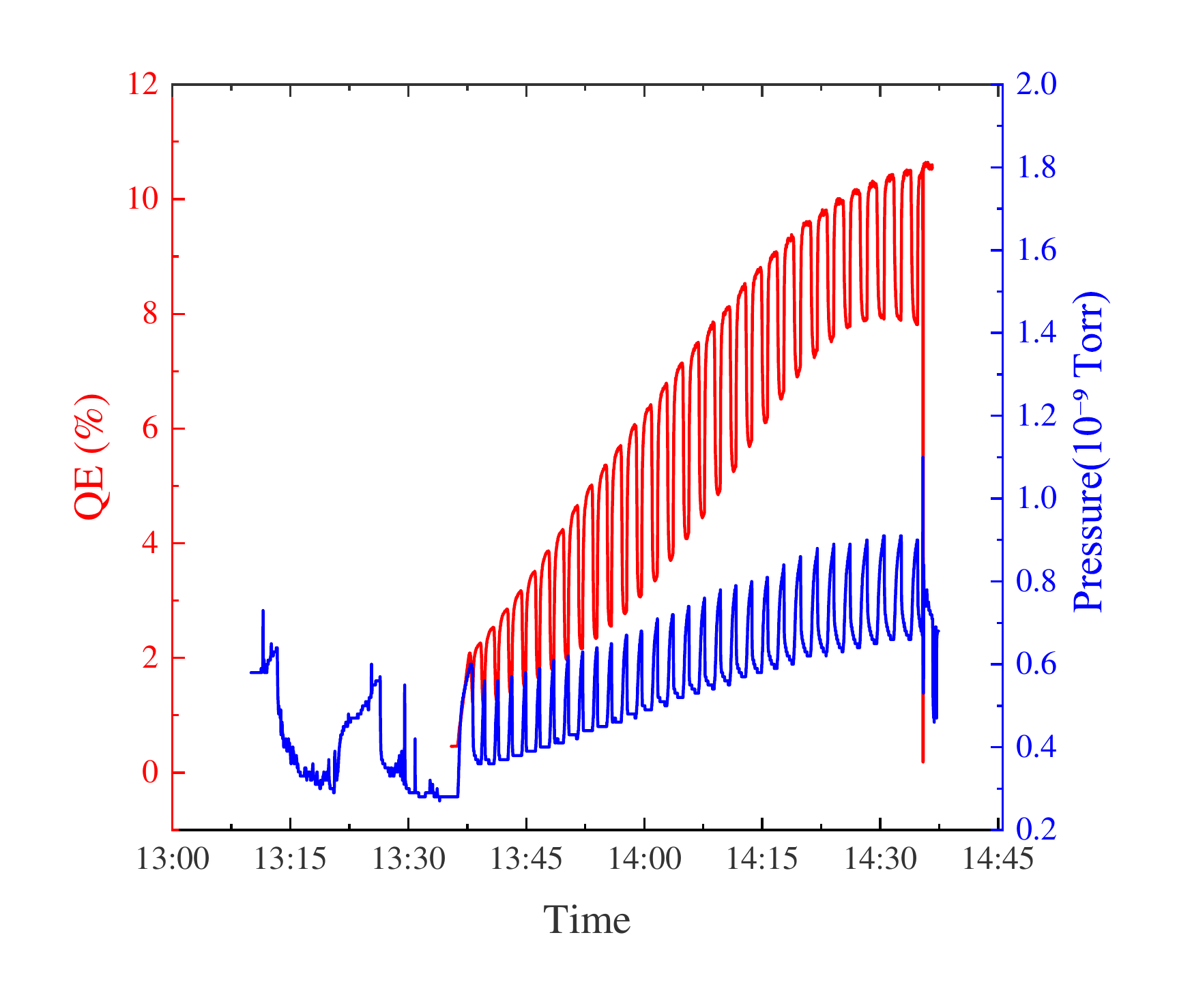}

            \put(17,46){%
                \includegraphics[
                    width=0.35\linewidth
                ]{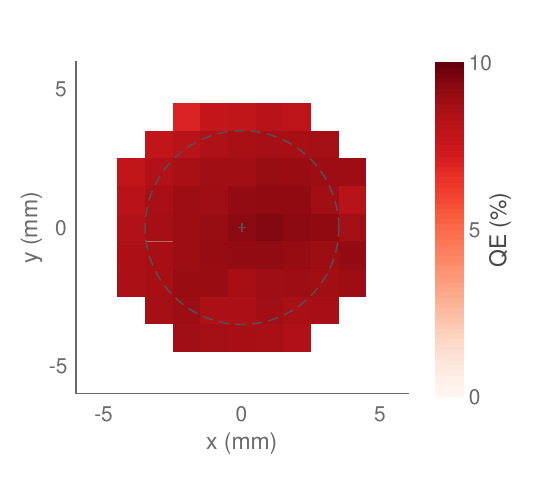}%
            }

        \end{overpic}

        \caption{K-rich K$_2$CsSb photocathode.}
        \label{fig:qe_pressure_rk}
    \end{subfigure}
    \caption{Preparation procedures and final quantum-efficiency distributions of conventional and K-rich K$_2$CsSb photocathodes. The red curves show the quantum efficiency (QE, left axis), while the blue curves show the preparation-chamber pressure (right axis). (a) Conventional K$_2$CsSb photocathode, exhibiting a maximum QE of 6.3\%, a mean QE of 5.6\% over the active area, and a uniformity of 92.4\% 
    within a 7-mm-diameter region. (b) K-rich K$_2$CsSb photocathode prepared with an extended K-growth stage prior to the common Cs-activation step. The final Cs-activation process is identical for both photocathodes. The K-rich photocathode exhibits a maximum QE of 9.5\%, a mean QE of 8.6\%, and a uniformity of 94.7\% within the same 7-mm-diameter region. The insets show the final two-dimensional QE distributions, with the coordinates referenced to the fitted photocathode centers. The dashed circles indicate the 7-mm-diameter regions used to evaluate uniformity. Both QE maps use the same 0--10\% color scale for direct comparison.}
    \label{fig:qe_pressure}
\end{figure}

In this work, an optimized growth recipe was developed for the fabrication of bialkali photocathodes.  Photocathodes prepared using both the conventional and optimized recipes were investigated to determine how the growth procedure affects cryogenic photoemission. The principal modification is an extended K-deposition stage that introduces a substantially larger K exposure during growth. Both conventional and K-rich photocathodes are subsequently subjected to the same YOYO Cs/Sb activation procedure until QE saturation. Fig.\ref{fig:qe_pressure} compares the preparation of Conventional and K-rich K-Cs-Sb photocathodes. 

\begin{figure}[!htbp]
    \centering
    \includegraphics[width=0.88\linewidth]{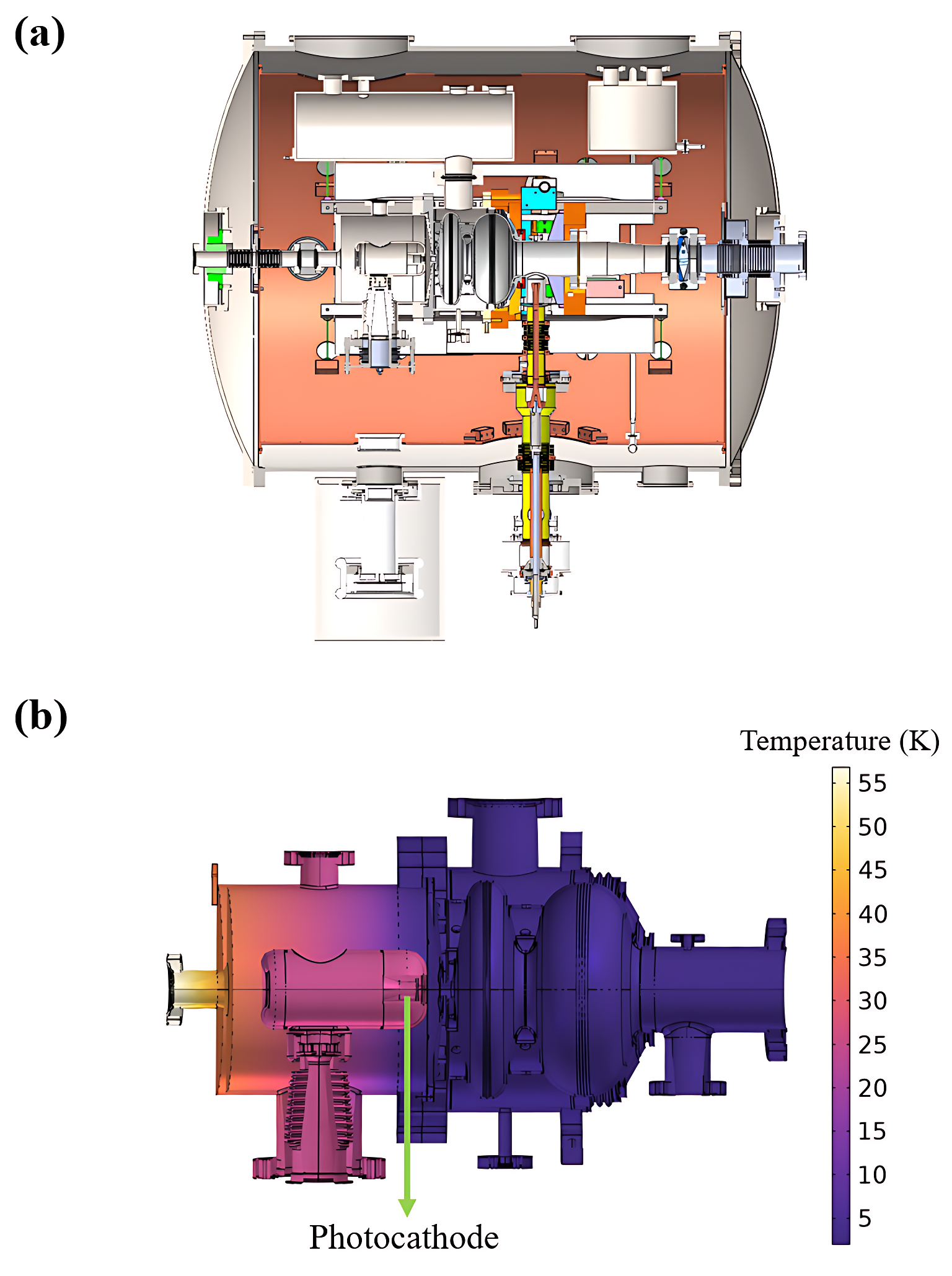}
    \caption{Cryogenic electron-gun configuration and cathode temperature. (a) Cross-sectional view of the DC-SRF electron gun, showing the photocathode insertion assembly. (b) Simulated temperature distribution in the cathode assembly and adjacent gun structures. The photocathode reaches a temperature of approximately 30 K when the SRF cavity is cooled with 2 K liquid helium.}
    \label{fig:dcsrf}
\end{figure}

After fabrication, the photocathodes were transported in an
ultrahigh-vacuum suitcase and transferred into the DC-SRF gun without
measurable degradation, consistent with our previous report~\cite{Ouyang2022BialkaliDCSRF}. The photocathode was transferred under ultrahigh vacuum(10\textsuperscript{-9} Pa) from the suitcase to the DC-SRF gun. As shown in Fig.\ref{fig:dcsrf},  all the QE and operation lifetime of the photocathode in the following paper were evaluated under realistic XFEL or high average current operating conditions in the gun rather than in an isolated laboratory test chamber. Fig.~\ref{fig:dcsrf}(b) shows that the photocathode is located in the cathode ball conduction cooled by the 2 K liquid helium which is used to cool the SRF cavity. The calculated cathode temperature is approximately 30~K and the vacuum of the photocathode is in low 10\textsuperscript{-9} Pa scale. 
\begin{figure}[t]
    \centering
    \includegraphics[width=0.8\linewidth]{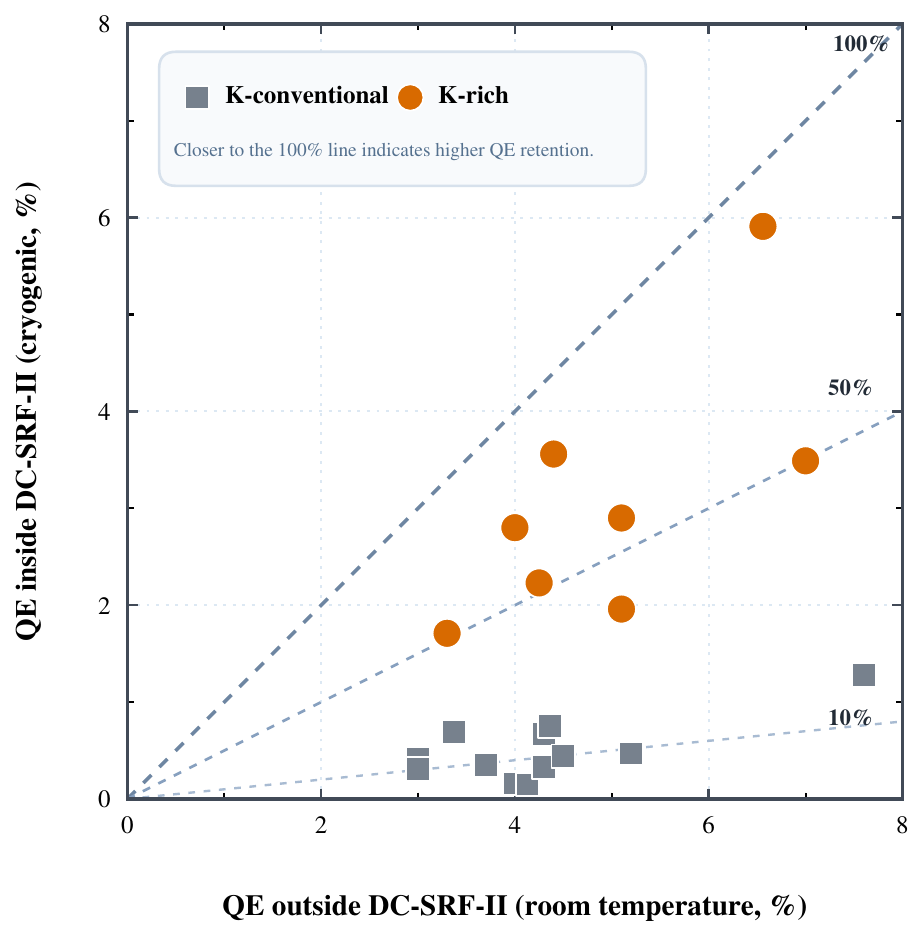} 
    \caption{Quantum efficiency (QE) measured in the suitcase at room temperature versus that measured after transfer and cooling to cryogenic temperature. Each point represents an independently prepared photocathode. Gray squares and orange circles represent Conventional (n=12) and K-rich  photocathodes (n=8), respectively. The dashed lines indicate constant QE-retention levels of 100\%, 50\%, and 10\%, where QE retention is defined as $\mathrm{QE}_{Cryo}/\mathrm{QE}_{\mathrm{RT}}\times 100\%$. Overall, the K-rich photocathodes exhibit higher QE and significantly improved QE retention after transfer and cooling.  }
    \label{fig:qe}
\end{figure}

The QE evolution from room temperature to cryogenic operation further highlights the critical role of photocathode growth procedure in maintaining efficient photoemission at low temperature. While Conventional recipe photocathodes exhibit a substantial loss of QE after cooling, the K-rich photocathodes preserve a significantly larger fraction of their initial QE under the same cryogenic conditions as shown in Fig.\ref{fig:qe}. Since the cooling procedure and accelerator environment were essentially identical for the two groups, including the rapid insertion of the photocathodes into the precooled gun, the markedly different QE retention cannot be explained by the cooling procedure alone. The present results therefore indicate that fast cooling alone is insufficient to determine the cryogenic QE; the photocathode preparation history provides an additional controlling variable. This observation suggests that material preparation can compensate for part of the cryogenic QE loss without increasing the transverse energy of the emitted electrons.

\begin{figure}[t]
    \centering
    \includegraphics[width=0.82\linewidth]{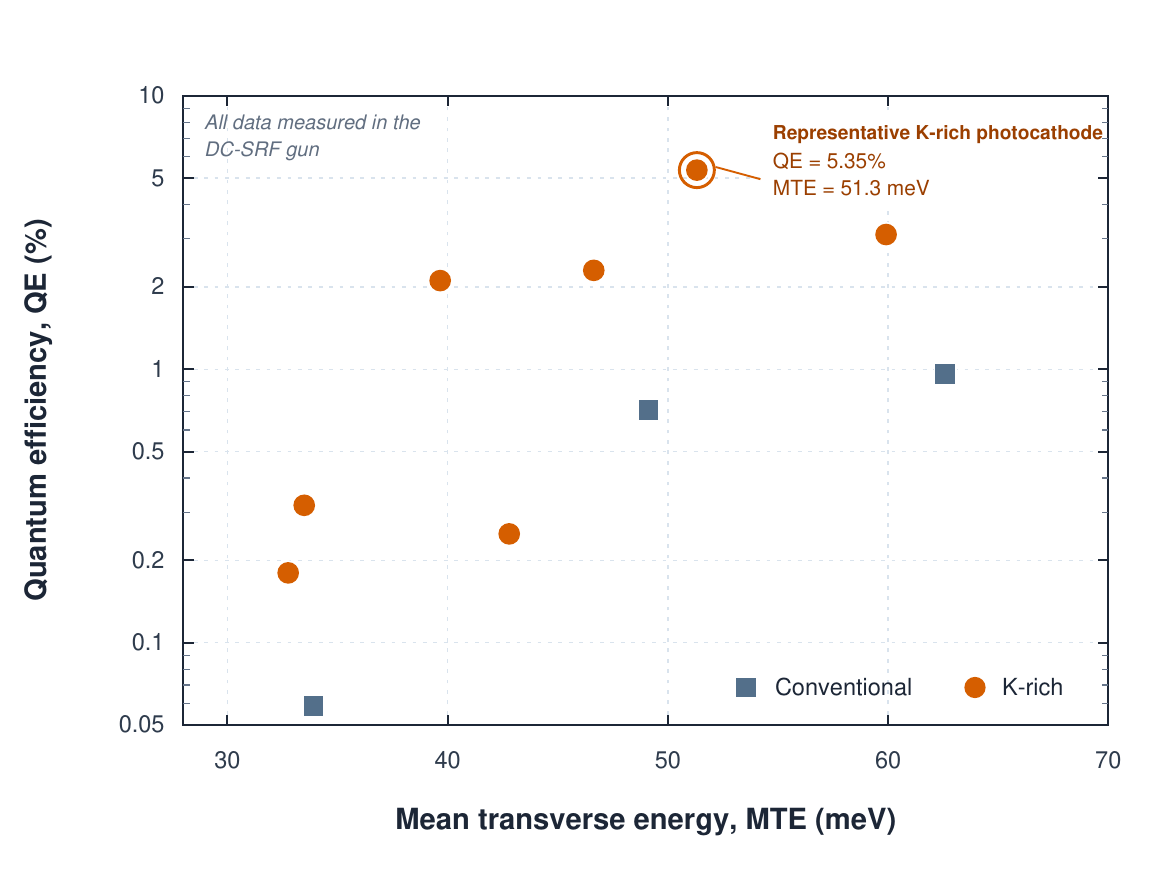}
    \caption{QE–MTE performance of cryogenic K$_2$CsSb photocathodes during beam experiments. Each marker corresponds to an independently prepared photocathode measured under same cryogenic conditions. The QE axis is logarithmic. }
    \label{fig:qe_emittance}
\end{figure}

 The MTE of the bialkali photocathode is measured  using slit scan method. The detailed measurement method can be found in reference\cite{HuangERL2026}. In
Fig.\ref{fig:qe_emittance},  the MTE of bialkali photocathodes prepared using the Conventional recipe and the optimized K-rich YOYO recipe is compared. In both cases, the normalized emittance  exhibits an excellent linear dependence on the rms laser spot size, confirming that the measured MTE is dominated by the intrinsic photoemission process. Linear fits yield intrinsic-emittance slopes corresponding to MTE values of approximately 50 meV for the Conventional recipe and K-rich photocathodes. Compared with the room-temperature MTE value of approximately 160 meV\cite{Bazarov2011CsK2SbThermalEmittance,Xu2024ThermalEmittanceBialkaliPKU}, cryogenic operation reduces MTE by more than 65\%, providing a substantial increase in the achievable beam brightness. Remarkably, the optimized K-rich recipe preserves the low MTE while increasing the QE by more than a factor of five. The optimized growth procedure substantially increases cryogenic QE while preserving the ultralow MTE, overcoming the conventional expectation that improved photoemission efficiency must be accompanied by increased mean transverse energy. 

Notably, the K-rich photocathodes also access an even lower-MTE operating regime. The lowest measured MTE is 32.8 meV with a QE of 0.18\%, while measurements in the 30–40 meV range retain QE in the 0.1-0.5\% range.  To our knowledge, an MTE of 32.8~meV represents one of the lowest values obtained from a semiconductor photocathode directly characterized in an operational electron-gun relevant to MHz-repetition-rate XFEL sources. Together with the 5.4\%-QE, 51.3-meV operating point, these measurements reveal a broad accessible QE--MTE performance space rather than a single optimized operating condition.

\begin{figure*}[htbp]
    \centering

    \begin{minipage}[t]{0.58\textwidth}
        \vspace{0pt}
        \centering
        \includegraphics[width=\linewidth]
        {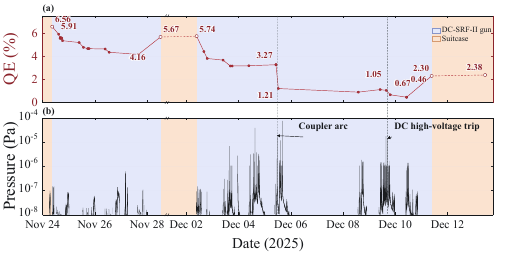}
    \end{minipage}
    \hfill
    \begin{minipage}[t]{0.40\textwidth}
        \vspace{0pt}
        \centering
        \includegraphics[width=\linewidth]
        {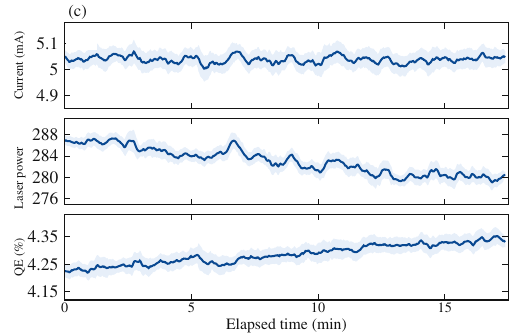}
    \end{minipage}

    \vspace{-1.5mm}

    \caption{
Long-term operational stability of the cryogenic
$\mathrm{K_2CsSb}$ photocathode.
(a) Evolution of the photocathode quantum efficiency over an
approximately 20-day operational history. The light-blue and
light-orange shaded regions indicate periods in the DC-SRF-II gun
and the ultrahigh-vacuum suitcase, respectively.
(b) Corresponding vacuum-pressure history. The labeled events indicate
major accelerator or vacuum interruptions.
(c) Photocurrent, incident laser power, and corresponding quantum
efficiency during 5.0-mA continuous-wave operation at a repetition
rate of 81.25 MHz. The mean photocurrent and quantum efficiency are
5.0 mA and 4.3\%, respectively.
    }
    \label{fig:operational_stability}
\end{figure*}

Fig.\ref{fig:operational_stability}(a) shows the QE evolution, while Fig.\ref{fig:operational_stability}(b) records the corresponding vacuum pressure.  The QE evolution exhibits two distinct contributions. First, during the initial cryogenic operation, the QE gradually decreased from approximately 6.6\% to 4.2-5.0\% and remained within this range for 5 days, but recovered to 5.7\% after the photocathode was returned to room temperature, indicating that a substantial part of the apparent cryogenic QE reduction was reversible. Second loss of QE was observed during the pressure-active period caused by major accelerator and vacuum events . The approximately 20-day operational history demonstrates extended usability under realistic machine conditions .  A QE of $4.3\%$ of the photocathode during 5 mA CW operation (81.25 MHz) is shown in Fig.~\ref{fig:operational_stability}(c), showing high QE photocathode survives high average current extraction. The detailed description of photocathode inside of the DC-SRF gun and beam experiments(1 MHz @ 100 pC) can be found in reference\cite{Jia2024HighBrightnessDCSRF,Li2025DarkCurrentDCSRF}. The bialkali photocathode with optimized K-rich YOYO recipe  demonstrates operational robustness relevant to CW XFEL injector conditions. 

 Previous cryogenic studies of alkali-antimonide photocathodes have shown that cooling can reduce MTE by reducing the excess emission energy, but often at the cost of reduced QE\cite{Xie2016CryogenicBialkaliPRAB,Mamun2017CryogenicTemperature,Owusu2025Cs3SbCryogenic}.
The comparison between photocathodes prepared by different growth procedures provides an empirical indication into the microscopic origin of the QE–MTE relationship. Although photocathodes made by two different recipes exhibit nearly identical MTE values under cryogenic operation, their QEs differ by more than an order of magnitude.  Within Spicer's three-step photoemission model, the QE is determined by the probabilities of photon absorption, carrier transport toward the emitting surface, and escape through the surface barrier, whereas the MTE is determined by the energy and angular distribution of the electrons which successfully escape\cite{Spicer1958Photoemission,Spicer1977Photoemission,Dowell2009Photocathode}. The big difference indicates the growth procedure strongly modifies the number of carriers that ultimately contribute to photoemission, but does not measurably broaden the transverse-energy distribution of the emitted electrons. This suggests that the enhancement arises from changes in carrier generation and transport rather than from an increase in the initial excess energy of the emitted electrons. 

Previous structural studies have shown that photocathodes nominally described as K$_2$CsSb can occupy different regions of the Cs--K--Sb compositional phase space. Gaowei \cite{Gaowei2017SputteredBialkali} observed by in-situ XRF and XRD that additional Cs deposition could drive a K$_2$CsSb-like layer toward KCs$_2$Sb, accompanied by changes in lattice structure and optical response; the resulting photocathodes still exhibited approximately 3.3\% QE at 530 nm\cite{Gaowei2017SputteredBialkali}. Similarly, synchrotron characterization of photocathodes\cite{Gaowei2019LEReCDC} prepared for the BNL LEReC gun showed atomic ratios close to KCs$_2$Sb rather than ideal K$_2$CsSb, demonstrating that the final photoemissive phase can differ substantially from the nominal material designation. 

Changes in the K/Cs ratio may alter the electronic structure, as first-principles calculations predict different band-edge characteristics for ideal crystalline K$_2$CsSb and KCs$_2$Sb. K$_2$CsSb is predicted to have a direct band gap at the $\Gamma$ point, whereas KCs$_2$Sb exhibits an indirect band gap involving electronic states at different points in the Brillouin zone~\cite{Ettema2002,Cocchi2019}. Thus, changes in the relative K/Cs content can modify not only the band-gap energy but also the momentum-space pathways involved in optical excitation.

This direct--indirect distinction provides one possible connection between growth and cryogenic QE. Direct optical transitions can conserve crystal momentum without participation of a lattice vibration, whereas indirect transitions require phonon-assisted momentum transfer. The phonon-assisted transitions depends explicitly on the phonon occupation, which decreases strongly upon cooling\cite{Zacharias2015,Noffsinger2012}  . A growth-induced modification of the relative contributions of K$_2$CsSb-like and KCs$_2$Sb-like electronic structures could therefore alter the sensitivity of the excitation process to cryogenic cooling. In particular, the contribution of phonon assisted channels is expected to become more strongly suppressed upon cooling, whereas direct optical excitation is comparatively less sensitive to phonon depletion.

Phonons also influence the subsequent transport step of photoemission. Cooling modifies electron--phonon scattering rates, energy relaxation times and carrier mean free paths, thereby changing the fraction of photoelectrons that can reach the emitting surface with sufficient energy to escape. The combined influence of phonon-assisted excitation and temperature-dependent carrier relaxation provides a natural route by which different Cs--K--Sb compositions can exhibit substantially different QE retention under otherwise similar cryogenic conditions. 

The nearly unchanged MTE provides an independent constraint on this interpretation. If the enhanced QE originated primarily from a substantial increase in photoelectron excess energy, an increase in transverse momentum spread would generally be expected. Experimentally, however, both preparation routes reach nearly the same low-MTE regime despite their very different QEs. The K-rich growth procedure therefore increases the fraction of carriers contributing to emission without measurably increasing the transverse energy of the emitted population. This observation is consistent with a picture in which the growth procedure mainly modifies the efficiency of carrier generation and transport, while cryogenic operation continues to determine the low-energy distribution of the emitted electrons. 

In summary, we demonstrate that cryogenic K$_2$CsSb photocathodes can simultaneously deliver percent-level QE, ultralow MTE, and robust long-term CW operation in an XFEL and ERL-relevant gun. More generally, these results suggest that photocathode growth and cryogenic operation provide complementary control over different stages of the three-step photoemission process, particularly carrier transport and electron escape. These results show that the practical QE-MTE trade-off can be substantially mitigated by combining growth optimization with cryogenic operation, and provide a practical pathway for next-generation high-brightness CW electron sources for XFELs and energy-recovery linacs, particularly those based on SRF guns or high-voltage cryogenic DC guns.

Acknowledgment: This work was supported by the National Natural Science Foundation of China (Grant No. 12575163). The authors thank Erdong Wang of Brookhaven National Laboratory for insightful discussions on the physical interpretation of the experimental results.

\bibliographystyle{apsrev4-2} 
\bibliography{photocathode_low_mte_recent_refs}

\end{document}